\documentclass[fleqn,10pt]{wlscirep}
\usepackage[utf8]{inputenc}
\usepackage[T1]{fontenc}
\title{Cavityless self-organization of ultracold atoms due to the feedback-induced phase transition}

\author[1,*]{Denis A. Ivanov}
\author[1]{Tatiana Yu. Ivanova}
\author[2]{Santiago F. Caballero-Benitez}
\author[1,3]{Igor B. Mekhov}
\affil[1]{St.~Petersburg State University, Ulianovskaya 3, Petrodvorets, St. Petersburg, 198504, Russia}
\affil[2]{Instituto de Física, Universidad Nacional Autónoma de México, Ciudad de Mexico 04510, Mexico}
\affil[3]{University of Oxford, Department of Physics,
Clarendon Laboratory, Parks Road, Oxford OX1 3PU, UK}

\affil[*]{d.a.ivanov@spbu.ru}

\begin{abstract}
Feedback is a general idea of modifying system behaviour  depending on the measurement outcomes. It spreads from natural sciences, engineering, and artificial intelligence to contemporary classical and rock music. Recently, feedback has been suggested as a tool to induce phase transitions beyond the dissipative ones and tune their
universality class. Here, we propose and theoretically investigate a system possessing such a feedback-induced phase transition. The system contains a Bose-Einstein condensate placed in an optical potential with the depth that is feedback-controlled according to the intensity
of the Bragg-reflected probe light. We show that there is a critical value of the feedback gain where the uniform gas distribution loses its stability and the ordered periodic density distribution emerges. Due to the external feedback, the presence of a cavity is not necessary for this type of atomic self-organization. We analyze the
dynamics after a sudden change of the feedback control parameter. The feedback time constant is shown to determine the relaxation above the critical point. We show as well that the control algorithm with the derivative of the measured signal dramatically decreases the transient
time.
\end{abstract}
\begin{document}

\flushbottom
\maketitle
\thispagestyle{empty}

\section*{Introduction}

Feedback control is known to be a useful tool to modify dynamics of classical\cite{fb-class} as well as quantum systems\cite{wiseman-milburn-book}. There were many theoretical proposals\cite{raizen,wallentowitz,averbukh,tombesi,jacobs,kilin,Szigeti2009,Szigeti2010,Hush2013,Gambetta2004,optica} and several experimental realizations\cite{rempe,rempe2,morrow,bushev,JThompson} of quantum feedback control in optics and atomic physics. 

The feedback control of quantum systems is fundamentally different from that of classical systems mainly due to the inevitable measurement back-action\cite{belavkin1983}. In most cases this quantum effect does not pose sever limitations on the feasibility of feedback control, but it has to be taken into account designing quantum control algorithms\cite{tan2000}.

Recently the feedback loop has been suggested as a tool to control phase transitions in quantum systems\cite{mekhov-fpt} and, in particular, in many-body settings\cite{optica,MazzucchiPRA2016}. The feedback control of atomic self-organization has been recently demonstrated experimentally in\cite{esslinger2019}. Thus the new class of feedback phase transitions (FPT) has been introduced, which can have properties beyond dissipative phase transitions in open systems. The key advantage of FPT is its extremely high degree of flexibility and controllability, which allows for the manipulation of the critical point and critical exponents and, therefore, enables the tuning and control of the universality class of phase transitions\cite{mekhov-fpt}. 

In this report we apply the concept of FPT to a system based on Bragg light scattering from a Bose-Einstein condensate (BEC). Instead of the self-formed standing-wave potential\cite{ritsch-no-cavity,ketterle-no-cavity}, we propose to use an actively controlled optical lattice with the control based on the Bragg-reflected signal of a probe light. The system with active feedback provides much higher degree of control on the distribution of atoms and its evolution around the critical point in comparison to systems without feedback. Such improved controllability can assist in quantum simulations, based on ultracold quantum gases\cite{CaballeroPRA2016,CaballeroNJP2015,CaballeroNJP2016,PhysRevA.96.051602,mekhov3}.

Worth noting that similar mechanism of self-organization has been investigated for atoms coupled to an optical cavity\cite{so-bec1,so-bec2,so-bec3,carl-therm,carl-bec1,carl-bec2,hemmerich,esslinger,esslinger17,esslinger17-2,lev17,Zimmermann2018, Ruostekoski2014, Morigi2010} [see for reviews\cite{rev-mekhov-ritsch,RitschRMP2013}]. There were also proposals to enhance the cavity self-organization by applying incoherent\cite{ivanov-jetp-lett,ivanov-jetp} and coherent\cite{Grimsmo_2014,Kopylov_2015} feedback. However the key advantage of the feedback – the free choice of the transfer function and its influence on the phase transition - was not discussed in these references. Various transfer functions can lead to the appearance of novel types\cite{mekhov-fpt} of time crystals\cite{Ueda2018,Zhu_2019,Angelakis2019} and Floquet engineering, as well as creation of quantum bath simulators\cite{mekhov-fpt} in many-body systems\cite{optica,MazzucchiPRA2016}. It will be intriguing to study, how more advanced methods than the
feedback control can influence quantum systems, for example, applying the digital methods of machine learning and artificial intelligence in real time.

\section*{Results and Discussions}

\subsection*{Model}
\label{sec:model}

\begin{figure}
\includegraphics[width=0.5\linewidth]{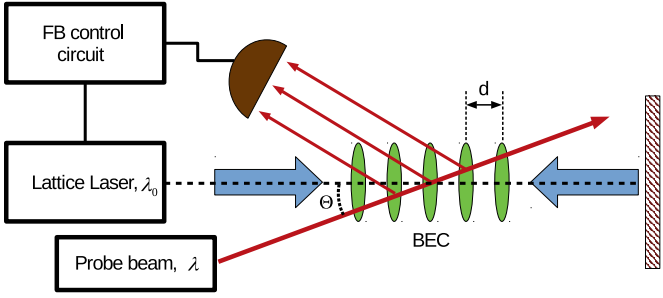}
\caption{The principle scheme of the setup. The BEC of atoms is trapped in a 1D optical lattice. The intensity of the lattice laser is controlled by the feedback loop. The measurement signal for the feedback loop is the light that is Bragg-reflected by the atomic sample.}
\label{fig:fig1}
\end{figure}

The considered FPT scheme can be applied to break the translational symmetry in a sample of any kind of polarizable particles including molecules~\cite{mekhov1-LP13}. To avoid the influence of thermal fluctuations we consider trapped atoms cooled below the Bose-Einstein condensation temperature. The ensemble is assumed to be sufficiently dilute so that the atom-atom interactions will be neglected. The atoms are placed into a one-dimensional (1D) optical lattice with the controlled potential depth, see Fig.~\ref{fig:fig1}. 

We restrict our consideration to 1D in order to focus on the most essential features of the feedback-induced dynamics. However, from the experimental point of view the realization of such a configuration requires additional potential that provides the confinement of the atoms along the axis of the lattice. The experimental realization of the proposed scheme might be more practical in 2D configuration\cite{esslinger2010}. Moreover, one should take into account possible stray feedback loops that can result in additional instabilities and pattern formations\cite{Ackemann1994,Tesio2014}. Here we would like to consider an idealistic situation and leave the mentioned important aspects for the future analysis. 

The feedback loop is organized as follows. The atoms are illuminated by weak but classical probe light directed at some angle $\Theta$ with respect to the axis of the optical lattice. The light reflected by the atoms is photodetected and the obtained signal is used to control the lattice potential.

The feedback algorithm is designed to increase the scattered light and provide tighter localization of the atoms near the minima of the lattice potential. This is achieved if the atomic density distribution fulfills the Bragg condition for the probe beam
\begin{equation}
\label{eq:bragg}
2 d \cos(\Theta) = \lambda.
\end{equation}
Here $d$ is the distance between the neighbouring peaks of the atomic density and $\lambda$ is the wavelength of the probe light. Taking into account that the distance between the potential minima in the lattice is $\lambda_0/2$, one can fulfill the Bragg condition for different atomic arrangements. Let us suppose that the wavelength of the probe beam is slightly smaller than that of the lattice potential. Than for small angle  $\cos{\Theta} \approx 1$ the Bragg condition will be satisfied for the atoms localized in each lattice well, which we will assume in this work.

It is important to mention one technical advantage of the feedback scheme shown in Fig.~\ref{fig:fig1} in comparison to systems without feedback. Since the operation is based on the reflection of additional probe light it is not necessary to use red-detunig with the atomic localization in strong-field regions.  The atoms can be localized in the anti-nodes of the blue-detuned lattice, while the information on their distribution can still be obtained by the reflection of the probe light.

An interesting generalizations of the measurement scheme might be possible. In particular, one can measure light scattered at an angle that does not satisfy the condition~(\ref{eq:bragg}), which would contain different information about the distribution of the atoms\cite{MekhovPRL2007,KozlowskiPRA2015,MekhovLasPhys2010,MekhovLasPhys2011}. The feedback-induced maximization of this signal might result in more exotic states of the atomic ensembles, both bosons and fermions\cite{MazzucchiNJP2016, MazzuchiSciRep2016, KozlowskiSciRep2017,CaballeroNJP2016,PhysRevA.96.051602,Schlawin-Jaksch, Colella_2019, Sheikhan2016, Fan2018, Mivehvar2017,Lev2019}. We leave this interesting possibility for future research.

\subsection*{Critical point}
\label{sec:crit_point}

We will describe the dynamics of BEC in a 3-mode approximation assuming that the lattice potential as well as the probe light excite only the lowest possible momentum eigenstates. The momentum kick due to the scattering of lattice photons is $2\hbar k_0$. Thus the atom field operator approximates as
\begin{equation}
\label{eq:3-mode-def}
\psi(x) = \frac{1}{\sqrt{L}} \psi_\mathrm{0} + \frac{1}{\sqrt{L}} \psi_\mathrm{L} \exp\left(2ik_0x\right) + \frac{1}{\sqrt{L}} \psi_\mathrm{R} \exp\left(-2ik_0x\right),
\end{equation}
where $L$ is the length of the BEC sample. The bosonic annihilation operators $\psi_\mathrm{0}$, $\psi_\mathrm{L}$, and $\psi_\mathrm{R}$ describe uniform distribution of the atoms, the first left- and right- running modes, respectively. 

To solve for the dynamics of the system we obtain and numerically simulate the semi-classical equations of motion for the average atomic fields $\alpha_\mathrm{0} \!=\! \langle \psi_\mathrm{0} \rangle$, $\alpha_\mathrm{L} \!=\! \langle \psi_\mathrm{L}\rangle$, $\alpha_\mathrm{R} \!=\! \langle\psi_\mathrm{R}\rangle$. In Methods it is shown that the equations of motion for these fields read 
\begin{eqnarray}
\label{eq:aver-evol}
\dot{\alpha}_\mathrm{0} &=& -i U_0 \eta\left(\alpha^* \alpha_\mathrm{R} + \alpha\alpha_\mathrm{L}\right) 
+ \frac{iU_\mathrm{0} I(t)}{2} \left(\alpha_\mathrm{R} + \alpha_\mathrm{L}\right), \nonumber\\
\dot{\alpha}_\mathrm{L} &=& -4i\alpha_\mathrm{L} - iU_0\eta\alpha^* \alpha_\mathrm{0} + \frac{i U_\mathrm{0} I(t)}{2} \alpha_\mathrm{0}, \nonumber\\
\dot{\alpha}_\mathrm{R} &=& -4i\alpha_\mathrm{R} - iU_0\eta\alpha \alpha_\mathrm{0} + \frac{i U_\mathrm{0} I(t)}{2} \alpha_\mathrm{0}.
\end{eqnarray}
Here we use $\eta$ for the probe field amplitude, $U_0$ for the atom field coupling, $I(t)$ for the intensity of the lattice field and $\alpha$ for the amplitude of the Bragg-scattered light. To simplify the notation and the presentation of the numerical results we scale time by the atomic recoil period $T_\mathrm{R}\!=\!2\pi/\Omega_\mathrm{R} \!=\! 4\pi m/\hbar k_0^2$ with respect to the lattice photons. The adiabatic scattered field amplitude is obtained averaging the quantum result Eq.~(\ref{eq:adiabatic_a}) in Methods: $\alpha = -2iU_\mathrm{0}\eta(\alpha_\mathrm{L}^*\alpha_0 + \alpha_0^*\alpha_\mathrm{R})/\kappa$. In order to simplify the analysis we assume that the scattered light is collected in a single mode of an auxiliary ring cavity that is not shown in Fig.~\ref{fig:fig1}. The photon decay rate $\kappa$ of this cavity is assumed to be so large that the system with such cavity becomes essentially equivalent to the system without a cavity at all.  

We set the lattice intensity to be determined by the control algorithm
\begin{equation}
    \label{eq:I-algorithm}
    I(t) = K s_\mathrm{\tau} = K \kappa\int_{0}^{t} e^{-\frac{t-t'}{\tau}} |\alpha(t')|^2 dt',
\end{equation}
where $\kappa$ plays the role of the photon detection rate. The parameter $\tau$ characterizes the response time of the feedback loop. Finally, $K$ is the feedback gain parameter. Eq.~(\ref{eq:I-algorithm}) is the result of averaging of the quantum equation~(\ref{eq:u-def}) obtained in Methods.

The exponential kernel in Eq.~(\ref{eq:I-algorithm}) is not the only possibility. It is used here to demonstrate some of the features of the active feedback control and can be substituted with another transfer function to result in important and non-trivial effects\cite{mekhov-fpt}.

It turns out that the critical point is determined by the combination of the feedback parameters $F = K\tau$. The critical value $F_\mathrm{c}$ reads 
\begin{equation}
\label{eq:real-condition}
 F_\mathrm{c}  = \frac{\sqrt{2} \kappa}{U_\mathrm{0}^3 \eta^2 N^2} \left(1 + \sqrt{1 + \left(\frac{U_\mathrm{0}^2 \eta^2 N}{\kappa}\right)^2}\right).
\end{equation}

Below this value only uniform (symmetric) distribution is possible with no light reflection from BEC. For $F > F_\mathrm{c}$ non-zero reflection will take place together with the broken translation symmetry of the atomic distribution. Interestingly, the critical feedback parameter $F_c$ scales as $1/N$ for large number of atoms $N$. Thus even very weak feedback can result in FPT if BEC is sufficiently large.

\begin{figure}
	\includegraphics[width=0.5\linewidth]{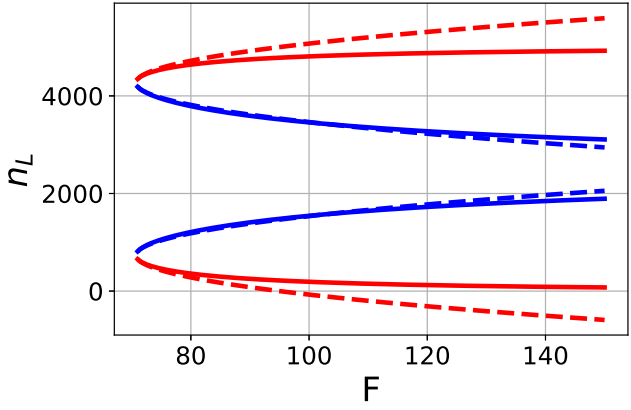}
	\caption{The dependence of the number of atoms in the left-running mode on the combined feedback parameter $F$ above the critical point. Dashed curves are the square-root approximations~(\ref{eq:n-delta}) in the vicinity of the critical point. For each $F>F_\mathrm{c}$ there are four stationary solutions. The uppermost and the lowermost branches (colored in red) correspond to unstable solutions and will not realize in an experiment. The parameters are: $\kappa \!=\! 2500$, $U_\mathrm{0} \!=\! 0.01$, $\eta \!=\! 1$ and $N \!=\! 10^4$.}
	\label{fig:fig2}
\end{figure}

It is remarkable that the stable solutions of the nonlinear system~(\ref{eq:aver-evol}) above the critical point can be analytically found. In particular one can find the steady-state values of the occupation numbers in different modes on the combined feedback parameter $F$. Below the critical point $F_\mathrm{c}$ (Eq.~\ref{eq:real-condition}) only uniform atomic distribution with zero average scattered field is possible. Above the critical point there are four different non-trivial stationary solutions. The results for one of the modes, $n_\mathrm{L}$, are shown in Fig.~\ref{fig:fig2} with solid lines. The uppermost and the lowermost branches (shown in red) are numerically found to be unstable. They will collapse to the uniform density distribution. Which of two stable branch will be realized in a particular experiment depends on the initial values and the noise in the feedback loop.

Varying the feedback parameter $F$ one can redistribute the atoms between different modes. In other terms this makes possible to control the depth of the atomic density modulation.   
   
The FPT above the critical point can be characterized by the functional dependence of atom numbers on the deviation from the critical point $\delta \!=\! F \!-\! F_\mathrm{c}$. The analytical result for small $\delta$ approximates as
\begin{eqnarray}
	\label{eq:n-delta}
	n_\mathrm{0} &=& n_\mathrm{0c} \pm 2U_\mathrm{0}N^2 \frac{\sqrt{U_\mathrm{0}^2 N^2 F_\mathrm{c}^2 + 2}}{\left(U_\mathrm{0}^2 N^2 F_\mathrm{c}^2 - 2\right)^2}  \sqrt{F_\mathrm{c} \delta}, \nonumber\\
	n_\mathrm{L} &=& n_\mathrm{Lc} \pm 2U_\mathrm{0}N^2 \frac{\sqrt{U_\mathrm{0}^2 N^2 F_\mathrm{c}^2 + 2}}{\left(U_\mathrm{0}^2 N^2 F_\mathrm{c}^2 - 2\right)^2} 
	\sqrt{F_\mathrm{c}\delta}, \nonumber\\
	n_\mathrm{R}  &=& n_\mathrm{L}, 
\end{eqnarray}
where $n_\mathrm{0c}$ and $n_\mathrm{Lc}$ are the critical values of the number of atoms in the zero-momentum and the left-running modes, respectively. The corresponding square-root dependencies are shown in Fig.~\ref{fig:fig2} with dashed curves. The stable solutions that are represented with two middle branches are very well approximated by the square-root dependencies of Eqs.~(\ref{eq:n-delta}).   

\subsection*{Transient dynamics}
\label{sec:tran_dynamics}

\begin{figure}
	\includegraphics[width=0.5\linewidth]{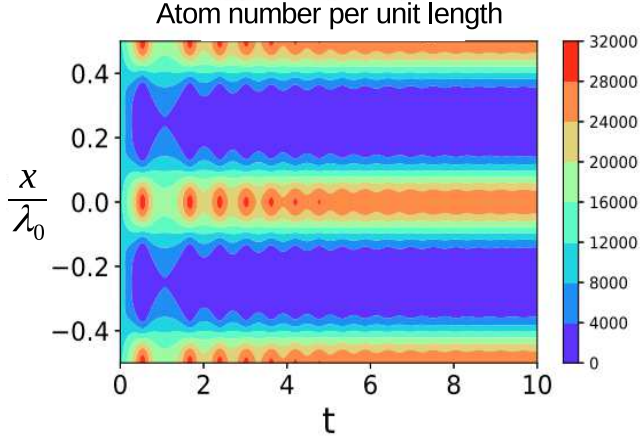}
	\caption{The time evolution of the atom distribution after a sudden switch on of the feedback control. The feedback strength $K \!=\! 6000$, time constant $\tau = 0.02$. The uniform distribution rapidly transforms into a lattice with the period $\lambda_0/2$. The position is measured in the units of the lattice laser wavelength $\lambda_0$.}
	\label{fig:fig3}
\end{figure}

To demonstrate the dynamics of the transition to the periodic atomic distribution from the uniform BEC we solve numerically Eq.~(\ref{eq:aver-evol}) assuming the following values of the parameters: $\kappa \!=\! 2500$, $U_\mathrm{0} \!=\! 0.01$, $\eta \!=\! 1$ and $N \!=\! 10^4$. These parameters are used in all of the considered examples. If we assume 87Rb atoms with $\Omega_\mathrm{R} \!=\! 2\pi \times 3,7$ kHz, then the used parameters correspond to collective coupling $U_\mathrm{0}N \!=\! 2\pi \times 0,37$ MHz and decay $\kappa \!=\! 2\pi \times 9,25$ MHz.  These values are quite moderate since in recent experiments much stronger coupling and smaller decay can be obtained~\cite{hemmerich,esslinger}. The feedback strength is $K \!=\! 6000$ and the feedback response time is $\tau \!=\! 0.02$ providing the parameter $F \!=\! 120$ to be higher then the critical one $F_\mathrm{c} \!=\! 70$. The initial condition for the feedback signal is set to a small non-zero value representing some noise that pushes the system from the unstable uniform distribution. The time evolution of the density distribution of the atoms inside three neighboring lattice sites is shown in Fig.~\ref{fig:fig3}.

The initially uniform distribution quickly transforms into the periodic pattern with the atoms grouped in the minima of the controlled optical potential. Thus above the critical point the atomic distribution looses its translation symmetry and the ordered phase emerges. 

The analytical solution of the system of the nonlinear stationary equations demonstrate that the combined feedback parameter $F$ determines the relative distribution of the atoms between the modes. However, it turns out that the values of $\tau$ and $K$ at fixed value of $F$ influence the system transient behavior. The numerical solution of Eqs.~(\ref{eq:aver-evol}) shows that the rate at which the steady state is reached strongly depends on $\tau$. 

\begin{figure}
	\includegraphics[width=1.0\linewidth]{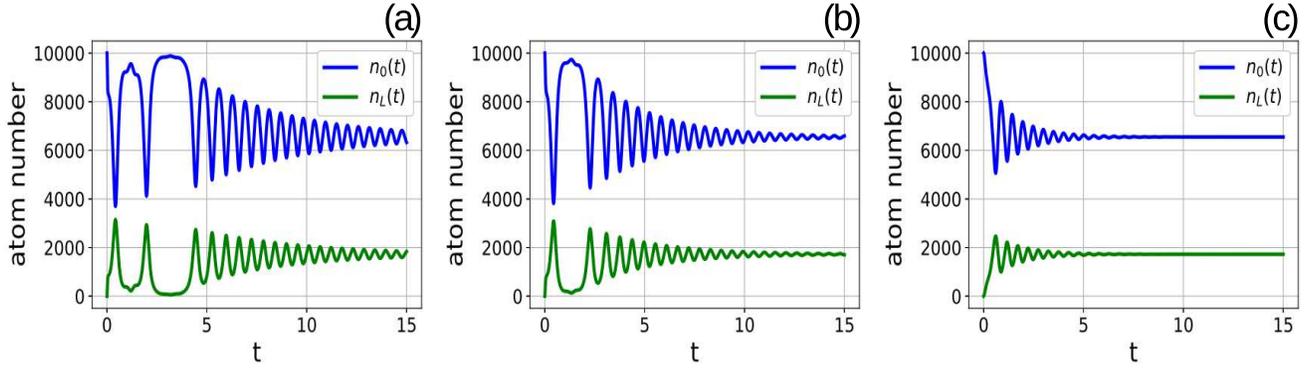}
	\caption{Numerical solutions of semi-classical equations for different values of the feedback response time $\tau$ and feedback strength $K$,  but the same value of the parameter $F = K\tau$. While the joint parameter $ F = K \tau$ is the same, the system behaviours are indeed different. The upper curve in each plot is the number of atoms in zero-momentum mode $n_\mathrm{0}(t)$, the lower curve is $n_\mathrm{L}(t)$. Feedback parameter above the threshold, $K \!=\! 9000$, response time $\tau \!=\! 0.013$ (subplot a), feedback parameter $K \!=\! 6000$, response time $\tau \!=\! 0.02$ (subplot b), feedback parameter  $K \!=\! 1200$, response time $\tau \!=\! 0.1$ (subplot c)}
	\label{fig:fig4}
\end{figure}

The time dependence of the number of atoms in the zero momentum (upper curves) and the left-running (lower curves) modes is shown in Fig.~\ref{fig:fig4}. The subplot $(a)$ corresponds to the feedback parameters above the critical value: $K \!=\! 9000$ and feedback response time $\tau \!=\! 0.013$. The results corresponding to the same feedback parameter $F$, but larger response times ($\tau \!=\!0.02$) are shown in the subplot $(b)$. The results for yet larger response time ($\tau \!=\! 0.1$) are shown in the subplot $(c)$. In these plots the upper curves represent the number of atoms in the zero-momentum mode, while the lower curves correspond to the left-running modes. In all tested cases there were $10^4$ atoms in the zero-momentum mode and no atoms in the excited modes. When the feedback starts at $t = 0$ the number of atoms in the zero momentum mode rapidly decreases with the simultaneous increase of the number of atom in the other modes. Then the oscillating transient behavior takes place until the numbers of atoms in different modes stabilize at their steady state values given in Eq.(\ref{eq:n-delta}). 

The time required for the system to reach the stationary regime depends on the value of the feedback response time $\tau$ even if the combined feedback parameter $F$ is fixed to provide the same steady state values. Comparing the subplots from $(a)$ to $(c)$ in Fig.~\ref{fig:fig4} one sees that the transient time decreases when the feedback time constant $\tau$ goes up. This indicates that the efficient strategy to transfer the atomic sample to the periodic pattern is to monotonically increase the potential. This emphasizes the important advantage of the electronic feedback over the cavity-based approach, since it is usually hard to make the cavity lifetime arbitrary long.

The electronic feedback allows for another improvement of the system performance in comparison to a system without feedback. The control $I(t)$ can contain not only the term proportional to the measured signal, but also, for example, its derivative:
\begin{equation}
\label{eq:pid-control}
I(t) = K s_\mathrm{\tau}(t) + K_\mathrm{d} \frac{d s_\mathrm{\tau}(t)}{dt}.
\end{equation}
In classical discussions of PID (proportional-integral-derivative) control~\cite{fb-class} it is shown that the derivative term can help to speed up the approach of the steady state, without change of the steady-state values. This feature is used for the feedback cooling of individual atoms~\cite{jacobs} and atomic ensembles at non-zero temperatures~\cite{jphysb,ivanov-jphb}. It turns out that the same effect takes place also for the considered here condensed atoms. The presence of the derivative in the control $I(t)$ is to certain extent equivalent to the presence of damping force that additionally reduces the oscillations and drives the system to the steady state. 

\begin{figure}
	\includegraphics[width=0.5\linewidth]{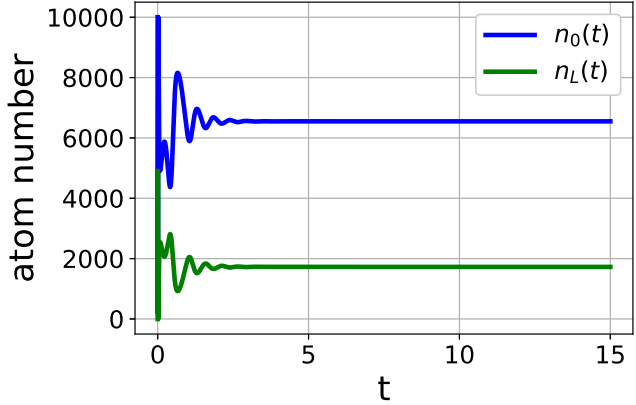}
	\caption{The evolution of the atom numbers in different modes for the feedback containing the derivative of the measured signal. The stabilization of the number of atoms is obtained much faster than for the similar case without derivative, compare with Fig.\ref{fig:fig4}. The feedback parameters are $K \!=\! 9000$, $\tau \!=\! 0.013$, $K_\mathrm{d} \!=\! -1000$.} 
	\label{fig:fig5}
\end{figure}

The effect of the derivative is demonstrated by numerical simulations. The example of the evolution of the system with the derivative in the control is shown in Fig.~\ref{fig:fig5}. The derivative feedback strength is $K_\mathrm{d} = -1000$. All other parameters are the same as in the subplot $(a)$ of Fig.~\ref{fig:fig4}. The comparison of Fig.~\ref{fig:fig4} and \ref{fig:fig5} clearly shows that the transient time is greatly reduced if the control action contains the derivative of the measured signal, as expected from the classical PID-control reasoning. Thus the appropriate feedback can be used to control not only the presence of the transition from the uniform to periodic density distribution, but also the rate of this transition which is impossible in a system without feedback.  

\section*{Conclusions}
\label{sec:conclusions}

We presented the FPT in the system with atomic BEC, where the probe light was Bragg-scattered from BEC and used to control the additional optical lattice potential for the atoms. At the critical point the atomic density distribution looses its translation symmetry and the periodic density pattern emerges. This pattern works as a Bragg grating resulting in strong reflected light above the critical point. For the exponentially decaying feedback transfer function the phase transition is determined by the product of the feedback gain and the feedback loop response time. The critical value of this product has been analytically found as well as the stationary solutions for atom numbers above the transition point. 

Above the threshold four different stationary ordered solutions have been found with only two of them being stable and corresponding to observable phases. The time required to reach the stationary values depends on the value of the feedback response time $\tau$. We have shown that for larger value of $\tau$ the stationary regime is obtained faster. Yet faster transition to the ordered phase can be obtained for the feedback signal containing the derivative of the measured photo-current as in the classical PID-controlled systems. These results demonstrate the advantages one can have with the feedback phase transition in atom-optical systems, which can lead to new types\cite{mekhov-fpt} of time crystals\cite{Ueda2018,Zhu_2019,Angelakis2019} and Floquet engineering, as well as creation of novel quantum bath simulators\cite{mekhov-fpt}, in particular, in many-body systems\cite{optica,MazzucchiPRA2016, Scanning-Microscope, Scanning-Microscope-2, Vasilyev2020}, as well as tuning the universality class of phase transitions\cite{mekhov-fpt}. It will be intriguing to study, how more advanced methods than the
feedback control can influence quantum systems, for example, applying the digital methods of machine learning and artificial intelligence in real time.

\section*{Methods}
\subsection*{Quantum feedback}

After standard steps (adiabatic elimination of the excited atomic state etc.) the following Hamiltonian for the atoms and the scattered light can be written~\cite{meystre}
\begin{eqnarray}
\label{eq:H0}
&&H_\mathrm{0} = \hbar \Delta a^\dagger a + \int dx \psi^\dagger(x)\Big[-\frac{\hbar^2}{2 m} \partial_x^2 
\\
&+& \hbar U_\mathrm{0} \left(a^\dagger a + |\eta|^2 + a^\dagger \eta e^{2ik_0x} + \eta^* a e^{-2ik_0x}\right)\Big]\psi(x), \nonumber
\end{eqnarray}
where $a$ is the annihilation operator of the light scattered from BEC, $\psi(x)$ is the atomic field operator obeying $\left[\psi(x),\psi^\dagger(x')\right] \!=\! \delta(x-x')$, $\Delta$ is the atom-field detuning, $U_\mathrm{0}$ is the atom-field interaction constant, $m$ is the mass of an atom. The probe field amplitude $\eta$ is assumed to be constant and real. This Hamiltonian contains interactions of the atoms with the probe and the scattered field only. The effect of the feedback controlled lattice potential and the measurement will be introduced later.   

Having the approximation~(\ref{eq:3-mode-def}) one obtains from Eq.~(\ref{eq:H0}) the following Hamiltonian 
\begin{eqnarray}
\label{eq:3-mode-H}
H_\mathrm{0} &=& 4\Omega_\mathrm{R} \left(\psi_\mathrm{R}^\dagger \psi_\mathrm{R} + \psi_\mathrm{L}^\dagger \psi_\mathrm{L}\right) \nonumber \\
&+& U_\mathrm{0}\eta \left[a^\dagger \left(\psi_\mathrm{0}^\dagger \psi_\mathrm{R}  + \psi_\mathrm{L}^\dagger \psi_\mathrm{0}\right) + a \left(\psi_\mathrm{0}^\dagger \psi_\mathrm{L}  + \psi_\mathrm{R}^\dagger \psi_\mathrm{0} \right) \right]
\end{eqnarray}
Here we also defined the recoil frequency $\Omega_\mathrm{R} \!=\! \hbar k_0^2/2m$. For simplicity it is assumed that $ \Delta \!=\! 0 $.

For simplicity we represent the reflected field as a running mode of an auxiliary cavity with large photon decay rate $\kappa$. In this case the field $a$ adiabatically follows the dynamics of the matter. Thus neglecting its time derivative in the Heisenberg equation for $a$ derived from the Hamiltonian Eq.~(\ref{eq:3-mode-H}) one finds the field as
\begin{equation}
\label{eq:adiabatic_a}
a \approx -\frac{2i U_\mathrm{0}\eta}{\kappa}\left(\psi_\mathrm{0}^\dagger \psi_\mathrm{R} + \psi_\mathrm{L}^\dagger \psi_\mathrm{0}\right).
\end{equation}

The measured quantity is the flux of the Bragg-reflected photons. Assuming ideal quantum efficiency of the detector and its wide bandwidth the conditioned evolution of the quantum state reads~\cite{wiseman-milburn-book}
\begin{equation}
\label{eq:me-meas}
d \rho_\mathrm{m}(t) = \left\{dN(t)\mathcal{G}\left[c\right] - dt \mathcal{H}\left[i\left(H_0 + H_\mathrm{fb}\right)+ \frac{1}{2} c^\dagger c\right] \right\}.
\end{equation}
The jump operator $c\!=\!\sqrt{\kappa} a$ contains the field annihilation $a$ that is expressed via atomic operators as given by Eq.~(\ref{eq:adiabatic_a}). After the adiabatic elimination of the scattered field $a$ the jump operator contains only the atomic operators. The increment of the Poisson stochastic process $dN(t)$ equals ether $0$ or $1$ such that 
\begin{equation}
\label{eq:EdN}
E[dN]\!=\!\mathrm{Tr}\left\{c^\dagger c \rho \right\}dt.
\end{equation}
The superoperators $\mathcal{G}$ and $\mathcal{H}$ read
\begin{eqnarray}
\label{eq:GH-def}
\mathcal{G}\left[A\right]\rho &=& \frac{A\rho A^\dagger}{\mathrm{Tr}\left\{A\rho A^\dagger \right\}} -\rho , \\
\mathcal{H}\left[A\right]\rho &=& A\rho + \rho A^\dagger - \mathrm{Tr}\left\{A\rho + \rho A^\dagger \right\} . \nonumber
\end{eqnarray}

The feedback action on the atoms is realized via the controlled standing wave potential with the coordinate dependence $\sim \cos(2k_0x)$. In the used 3-modes approximation the feedback Hamiltonian is given by
\begin{equation}
\label{eq:Hfb}
H_\mathrm{fb} = \frac{U_\mathrm{0} I(t)}{2} \left(\psi_\mathrm{0}^\dagger (\psi_\mathrm{R} + \psi_\mathrm{L}) + (\psi_\mathrm{R}^\dagger + \psi_\mathrm{L}^\dagger) \psi_\mathrm{0}\right) .
\end{equation}

Contrary to an instantaneous (Markovian) feedback, where the control depends on the measured outcome at the same time instant, $s_\tau (t)$, we consider a more general case where the finite response time of the feedback loop is taken into account
\begin{equation}
\label{eq:u-def}                        
I(t) = \mathrm{K} s_\mathrm{\tau}(t) = \mathrm{K} \int_0^t e^{-\frac{t-t'}{\tau}} dN(t') .
\end{equation}
For the aims of this report it is enough to take the exponential kernel with the feedback response time $\tau$ considered as one of the feedback parameters. Te other parameter is the feedback gain $K$. The effects emerging from the use of more general kernels are discussed in ~\cite{mekhov-fpt}. Here we address more technical aspects of the feedback operation and analyze the dynamics of the gas during the feedback control. Using semi-classical approach we test two feedback algorithms: proportional feedback and proportional with derivative.

\subsection*{Semi-classical steady-state solution}

To go to the semi-classical representation~(\ref{eq:aver-evol}) we calculate the evolution equations for the averaged amplitudes using Eq.~(\ref{eq:me-meas}) and neglect there quantum correlations between different bosonic modes of the system.

For the semi-classical representation of the measured photon number the stochastic increments are substituted with their expectation values $dN\!=\!\mathrm{Tr}\left\{c^\dagger c \rho \right\} dt$. 

In order to find the steady state solutions of Eqs.~(\ref{eq:aver-evol}) we represent the field amplitudes as $\alpha_\mathrm{0,L,R}(t) \!=\! \alpha_\mathrm{0,L,R} \exp\left(i\omega t\right)$. The eigenfrequency $\omega$ has to be found during the solution. After the substitution of this ansatz in Eqs~(\ref{eq:aver-evol}) one obtains the set of nonlinear algebraic equations that can be analytically solved. The stationary solutions for the atom numbers read
\begin{eqnarray}
\label{eq:SS-results}
n_\mathrm{0} &=& |\alpha_\mathrm{0}|^2 = \frac{N}{2} \frac{4+\omega}{2+\omega}, \nonumber\\
n_\mathrm{L} &=& |\alpha_\mathrm{L}|^2 = n_\mathrm{R} = |\alpha_\mathrm{R}|^2 = \frac{N}{4} \frac{\omega}{2+\omega}.
\end{eqnarray}
The eigenfrequencies can be found as real roots of the following 4-th order algebraic equation
\begin{equation}
\label{eq:omega-eq}
\left(4+\omega\right)\omega - \frac{2(2+\omega)^2}{U_\mathrm{0}^2 N^2 K^2\tau^2} \left(1 + \frac{\kappa^2 \left(2 + \omega\right)^2}{4 U_\mathrm{0}^4 \eta^4 N^2}\right) = 0. 
\end{equation}
The results in Eqs(\ref{eq:SS-results}) are obtained assuming that the stationary value of the feedback signal $I(t)$ is not equal to zero. Obviously, other solutions of Eqs~(\ref{eq:aver-evol}) are possible, where the feedback signal and the scattered light equal zeros. One trivial solution in this case is when all of the atoms are in the zero momentum mode, so that $\alpha_0 \!=\! \sqrt{N}$ and $\alpha_\mathrm{R} \!=\! \alpha_\mathrm{L} \!=\! 0$. This solution is unstable and will evolve after a small perturbation for the feedback below the critical point. Other solutions are with $\alpha_\mathrm{0} \!=\! 0$ and arbitrary distribution of all of $N$ atoms between left- and right-running modes. These solutions are stable. The characteristic equation~(\ref{eq:omega-eq}) is quadratic with respect to $(\omega \!+\! 2)^2$ and can be analytically solved. Analyzing this solution one can easily formulate the condition for $\omega$ to have zero imaginary parts, i.e. Eq.~(\ref{eq:real-condition}). The possible critical values of the eigenfrequencies are given by 
\begin{eqnarray}
	\label{eq:n-critical}
	\omega_\mathrm{c} = -2 \pm \frac{U_\mathrm{0}^2 N \eta^2}{\kappa}\sqrt{\left(U_\mathrm{0}^2 N^2 F_\mathrm{c}^2 -2 \right)} .
\end{eqnarray}
Inserting these results in Eq.(\ref{eq:SS-results}) one obtains the critical values of the occupation numbers for different modes: $n_\mathrm{0c}$, $n_\mathrm{Lc}$ and $n_\mathrm{Rc}$. 

\bibliography{my}

\section*{Acknowledgements (not compulsory)}

The financial support is provided by RSF 17-19-01097-P (St. Petersburg State University, AMO physics approach), DGAPA-UNAM IN109619 and CONACYT- 364 Mexico A1-S-30934 (Universidad Nacional Autónoma de México, CMT approach), and EPSRC EP/I004394/1. The research materials supporting this publication can be accessed by contacting Dr. Igor Mekhov at igor.b.mekhov@gmail.com.

\section*{Author contributions statement}

All the authors equally contributed to the results. I.B. and S.F. were more involved in analytic calculations, while D.A. and T. Yu. did more numeric simulations.


\end{document}